# Addressing Data Scarcity in UBEM Validation: Application of Survey Sampling Techniques

Chunxiao Wang[1, 2, 3], Bruno Duplessis[3], Eric Peirano[1], Pascal Schetelat[2], Peter Riederer[2]

[1]Efficacity, France

[2]CSTB (Scientific and Technical Center for Building), France

[3]CEEP (Energy, Environment and Processes), Mines Paris – PSL, France

## ABSTRACT

Urban Building Energy Models (UBEM) are vital for enhancing energy efficiency and sustainability in urban planning. However, data scarcity often challenges their validation, particularly the lack of hourly measured data and the variety of building samples. This study addresses this issue by applying bias adjustment techniques from survey research to improve UBEM validation robustness with incomplete measured data. Error estimation tests are conducted using various levels of missingness, and three bias adjustment methods are employed: multivariate imputation, cell weighting and raking weighting. Key findings indicate that using incomplete data in UBEM validation without adjustment is not advisable, while bias adjustment techniques significantly enhance the robustness of validation, providing more reliable model validity estimates. Cell weighting is preferable in this study due to its reliance on joint distributions of auxiliary variables.

## INTRODUCTION

Rapid urbanisation poses significant challenges in terms of energy consumption and sustainability, with urban buildings accounting for 36% of global energy use and 40% of $CO_2$ emissions. Urban Building Energy Models (UBEM) have emerged as a vital tool for enabling informed decision-making at the urban scale. Recent developments in UBEM have expanded their application to more intricate use cases, such as grid stability and load shifting studies (Ang et al., 2020). However, the effectiveness of these tools in such use cases depends heavily on their accuracy in predicting dynamic outputs, which underscores the importance of thorough model calibration and validation.

A critical practical issue in validating UBEM's dynamic outputs is data scarcity. Very few models capable of producing hourly (or even sub-hourly) results have been validated using hourly measured data (Lefort, 2022; Oraiopoulos & Howard, 2022). Furthermore, even when these measurements are available, they are often incomplete due to various practical issues, making missing data inevitable (Morewood, 2023). In many studies, high temporal resolution data are only collected over short periods, and practical issues with data collection and transmission can lead to significant data loss.

In the field of building energy modelling, researchers have attempted several strategies to mitigate the impact of missing data; however, these methods are inadequate for UBEM validation. Excluding buildings with missing data reduces an already limited number of validation experiments, diminishing their representativeness for the entire building stock. Limiting validation to shorter periods hinders the ability to generalise the model's validity to broader temporal contexts, such as different weather scenarios (Wang et al., 2024). Other solutions, such as linear interpolation and replacing missing values with data from subsequent years, are unsuitable due to their inability to accurately reconstruct missing high-resolution data and their disregard for temporal dynamics.

To address this research gap, the present study proposes adopting bias adjustment techniques from survey research to enhance the robustness of UBEM validation with incomplete measured data. These techniques are designed to make reliable inferences from incomplete datasets, thereby improving the internal validity of studies.

Section 2 outlines the general methodology, including the basic assumptions and techniques adopted in this study. Section 3 describes the measured data collection process and external data sources. Section 4 details the implementation of different methods and the design of the error estimation tests. Section 5 presents the results of error estimation tests using different ratios of missing data.

## METHODOLOGY

### Survey sampling analogy

In this study, we address the issue of data scarcity by treating it as a missing data problem commonly encountered in survey sampling. We draw the following analogy: measurements of heating loads are viewed as "responses" to a "survey," determined by physical parameters and boundary conditions incorporated into a building energy simulation. Incomplete measurements in some buildings or substations are considered as partial nonresponse. Table 1 illustrates the elements of UBEM validation and their equivalent concepts in survey sampling.

*Table 1: Elements of UBEM validation in an analogy of survey sampling*

| Concepts in Survey Sampling | Elements of Model Validation |
|---|---|
| Objective | Obtain measured data over the intended validation period |
| Population | Targeted building stock |
| Sample | Buildings with measured data |
| Individual | Building |
| Response | Measurements for each time step |
| Noncoverage | Buildings not sampled |
| Nonresponse | Sampled buildings without measured data |
| Partial nonresponse | Sampled buildings with incomplete measured data |
| Auxiliary data sources | Measured weather conditions, UBEM simulation results |

### Assumptions of missing data type

Depending on whether the systematic differences exist between missing data and available data, missing data can be categorized into three types: missing completely at random (**MCAR**), missing at random (**MAR**), and missing not at random (**MNAR**) (Mack et al., 2018).

- MCAR: If data is MCAR, it means the missingness is unrelated to any observed or unobserved data. For example, short, random interruptions in measurements can be considered as MCAR and do not bias the estimates.
- MAR: Data is classified as MAR when the likelihood of it being missing is related to the observed data, but not to the missing data itself. For example, if data often goes missing during cold periods, it can be considered MAR because the missing data correlates with outdoor temperature.
- MNAR: If the missingness is related to unobserved data, such as specific maintenance schedules or unknown system failures, it is termed MNAR, which can introduce significant bias.

The presence of MCAR data is considered as effectless, while for MAR and MNAR data, one must determine whether the missingness is independent of the outcome under study. In the context of UBEM validation, such missingness is not negligeable, as the under- or overrepresentation of specific weather conditions in the measured dataset can distort the calculated model validity. Therefore, we primarily focus on addressing long intermittences, which involve MAR and MNAR data, in our dataset. The effect of short intermittences (lasting from several hours to a day) is considered as MCAR and is thus ignored. Measurements with such MCAR missingness are used directly in model validation to calculate reference values of model validity.

*Table 2: Comparison across three methods on key aspects (MI: multivariate imputation, CW: cell weighting, RW: raking weighting. NC: not concerned; -: low; +: medium; ++: high)*

| | | MI | CW | RW |
|---|---|---|---|---|
| Requirements | Population | + | ++ | ++ |
| | Joint distribution | ++ | ++ | NC |
| | Marginal distribution | + | + | + |
| | Missing data structure | MAR in a class | MAR in a cell | MAR in a cell |
| Pros & cons | Nb of used variables | ++ | - | + |
| | Bias | + | - | + |
| | Variance | - | ++ | + |
| | Convergence | NC | NC | ++ |
| | Outliers | - | ++ | - |

### How to handle missing data

Partial nonresponse data may be handled by either weighting or imputation (Brick & Kalton, 1996). In this study, we apply multivariate imputation (**MI**), cell weighting (**CW**) and raking weighting (**RW**) to adjust the bias in estimated model validity caused by missing data (Kalton & Flores-Cervantes, 2003). These methods are qualitatively compared on some key aspects, as in Table 2, where the amount of knowledge required for each method is evaluated in the "Requirements" part.

In the current analogy, unfortunately the hypothesis of MAR data in a sub-sample (imputation class or cell) cannot entirely hold as we have little knowledge about the actual cause of missing data in real world. Nonetheless, for long intermittences in our dataset, we take the assumption of MAR to implement these bias adjustment methods. Besides, any efforts seeking to adjust the estimation bias should be beneficial compared to the ignorance of the missing data issue, which actually stands for a worst assumption – MCAR (Brick & Kalton, 1996).

## CASE STUDY

In this article, a validation with some incomplete measured data is performed for DIMOSIM[1], a UBEM developed by CSTB and Efficacity (Garreau et al., 2021). In this case study, we focus on the model capability to predict the heating demand of buildings.

In order to correct the bias caused by missingness, three types of data are mainly used: building consumption measurements, weather data and simulation results.

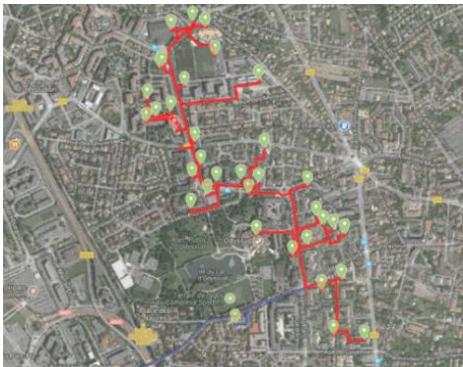

These datasets are detailed in the following sections.

Figure 1: Layout of district heating network in Blagnac

### Measurements

Measurements used in this study come from a district in Blagnac, France, where buildings are connected to a fourth-generation district heating network (Figure 1). The network features 4 kilometres of pipes, 36 substations and a nominal thermal power of 14 MW.

Measurements from heat meters installed at substations: average heating power per hour (MW), flow rate (m3/h), supply and return temperature (°C), with an hourly time step and monthly energy bills of every substation, are used in this study. All measured data have been cleaned and filtered to exclude mainly three types of data:

1. Unrealistic data, which are physically impossible,
2. Low-quality data (e.g., highly quantised signal),
3. Data during system malfunctioning, which cannot represent the actual demand patterns of buildings,
4. Inexplicable pattern data, which refer to unmanageable use scenarios.

The first type of data cleansing involves detecting outliers using the 3-sigma rule and performing internal verification between measures, such as comparing supply and return temperatures. The second type of data issue arises from the limited precision of heat meters and the relatively low flow rate during summer, which causes values to fall outside the functioning range of heat meters. Consequently, measurements which are highly quantised, as well as all summer hourly data where only DHW is consumed, are removed. For the third type of data cleansing, we apply a method proposed by Gadd and Werner to detect substation malfunctioning periods (Gadd & Werner, 2014). The fourth type of cleansing is unique to the context of model validation: it addresses periods or events where model validity is not the primary cause of inconsistence between measured and simulated data. For example, the validation errors during these periods can be due to the lack of knowledge about the actual usage scenario (maintenance or the start of heating season). Such lack of knowledge should not impact the validity or applicability of the model.

*Table 3: Features of sampled substations*

|       | Year of construction | Useful floor area (m2) | Missingness in 2021 |
|-------|----------------------|------------------------|---------------------|
| SST3  | 1976                 | 4907                   | 11%                 |
| SST8  | 1997                 | 4539                   | 14%                 |
| SST9  | 1976                 | 7567                   | 10%                 |
| SST13 | 1978                 | 3585                   | 12%                 |
| SST16 | 1979                 | 4949                   | 11%                 |
| SST20 | 1980                 | 6875                   | 13%                 |
| SST25 | 1987                 | 4827                   | 10%                 |

---

[1] DIMOSIM is the calculation engine of PowerDIS: https://efficacity.com/powerdis/

After preprocessing the data, we sampled buildings connected to seven residential substations for this study. These buildings were selected because they had relatively complete data for the validation period of the year 2021. The features and ratios of missingness of these substations are presented in Table 3. In these substations, the heating season end dates range from 20 to 28 May and the start dates range from 14 to 23 October. This results in a measurement period with at maximum from 5160 to 5352 hourly data points. The missing data points in these substations are considered as MCAR, as they are short intermittences randomly distributed over the measurement period, thus cause no bias on the model validity.

Weather data

Weather data of the city Blagnac are mainly measured at the airport of Toulouse-Blagnac, and solar radiation data is obtained from MERRA[2] and SODA[3]. Due to the close proximity of the meteorological measurement site to the district (only one kilometre), it was assumed that the meteorological data used would not introduce significant bias.

Given the fact that we only focus on the heating period of a year, all data from 29 May to 13 October are removed.

*Table 4: Calibrated parameters and their ranges*

| Name of parameter | Type | Range |
|---|---|---|
| user_draw_off_load (W) | rate | 0-2 |
| ExteriorWall_U_value (W/m$^2$K) | rate | 0.5-1.5 |
| ExteriorRoof_U_value (W/m$^2$K) | rate | 0.5-1.5 |
| ExteriorFloor_U_value (W/m$^2$K) | rate | 0.5-1.5 |
| ExteriorWall_window_U_value ((W/m$^2$K) | value | 1-6 |
| ExteriorWall_window_share (%) | rate | 0.5-1.5 |
| infiltration_rate (volume/hour) | rate | 0.5-1.5 |
| economy_heating_set_point (°C) | value | 16-19 |
| comfort_heating_set_point (°C) | value | 19-22 |
| economy_heating_start (hour) | value | 20-0 |
| comfort_heating_start (hour) | value | 4-8 |
| open_blind_ratio (%) | value | 0.2-0.9 |
| oversizing_coefficient (-) | value | 0.2-1 |

Simulation and calibration results

The sampled buildings were simulated using DIMOSIM, and the simulations were calibrated using Caliente—a calibration tool integrated with DIMOSIM, developed by CSTB. In this study, we adopted a straightforward calibration approach by generating a sample of model parameter combinations for DIMOSIM simulations, utilising Latin Hypercube Sampling (LHS). LHS was chosen for its ability to ensure that the sample captures the necessary variability effectively, even with a limited number of samples. Consequently, we selected a sample size of 200, constrained by computational resources. The calibrated model was then determined based on the parameter combination that resulted in the smallest validation error.

Due to the removal of hourly summer data, an initial stage of calibration was performed using monthly bills from the period without heating, which only included domestic hot water (DHW) usage. The second stage of calibration focused on other parameters, including the building envelope and system sizing coefficients. Table 4 summarises the ranges of all calibrated parameters, which are assumed to be uniformly distributed.

## IMPLEMENTATION

In this section, we present how the error generalisation test is designed and how bias-adjustment methods are implemented.

Error estimation test design

Using complete measured data, we first calculate the reference model validity for all sampled substations. The model validity is expressed through a simplified Goodness of Fit (**GOF**), which combines Normalized Mean Bias Error (**NMBE**) and Coefficient of Variation of the Root Mean Square Error (**CVRMSE**), as shown in Equation 1.

$$GOF = \frac{\sqrt{2}}{2}\sqrt{CVRMSE^2 + NMBE^2} \qquad (1)$$

To simulate data intermittences in measurements, we intentionally mask a certain percentage of the data. Since measurement interruptions generally lack discernible patterns, we assume that missing data occurs in continuous blocks, starting from a randomly selected date and hour. For each error generalisation test, we select 100 random start date-time indices from the validation period, and then continuously mask a specified percentage of measured data as missing. The percentage of missing data ranges from 5% to 95% to assess the usability of incomplete measured data under different levels of missingness.

---

[2] https://gmao.gsfc.nasa.gov/reanalysis/MERRA-2/

[3] https://www.soda-pro.com/

Three adjustment methods (imputation and weighting) are then implemented to estimate errors from the incomplete measured data. While comparing with reference errors, estimated errors may introduce bias and variance. That's why we investigate the following three indicators: Root Mean Squared Error (**RMSE**), median error and 95% confidence interval (**CI95**). The RMSE here, as defined in equation 2, is an error between the reference error calculated with complete measured data ($E_r$) and the estimated error ($E_e$) calculated with incomplete data. $n$ is the number of repetition tests for the same ratio of missingness.

As validations using hourly measured data target mostly calibrated models, only the best 5% parameter combinations with the lowest errors from the calibration are used in error estimation tests.

$$RMSE = \sqrt{\frac{\sum_{i=1}^{n}(E_r - E_{e,i})^2}{n}} \quad (2)$$

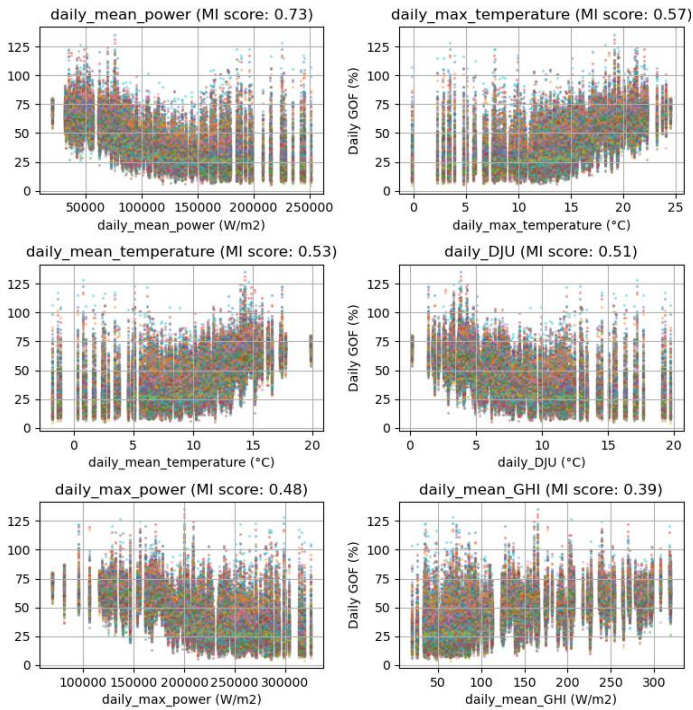

Figure 2: mutual information between daily GOF and extracted daily features.

### Sensitivity analysis
Before performing any bias adjustments, it is important to identify auxiliary variables from external data that can be useful. In this step, we conduct a sensitivity analysis to choose auxiliary variables from two major external data sources: weather data and simulation results.

Using the measurement data from substation 16, which was adopted due to the high quality of the measurements, and the simulation results of 200 parameter combinations from the calibration sample, validation errors were calculated with hourly time resolution and daily validation periods.

Then, daily features are extracted from both simulations and hourly weather data. These features are correlated with the daily GOF to calculate mutual information scores. The most related features from the simulation and weather data are presented in Figures 2, along with respective mutual information scores. These scores indicate the strength of the relationship between auxiliary variables and our variable of interest – validation error. Theoretically, daily validation errors can be also related to other aspects such as occupational behaviours inside buildings and control of heat supply in substations. However, such information is rarely directly available in most measurement campaigns, as it is in this study. Simulation results and measured weather data are therefore chosen as external data sources that we can use for imputation and weighting given their relationship with validation errors and data availability.

### Imputation adjustment
For each substation, we create a dataset with hourly auxiliary variables from weather data and an average simulated load profile from all simulations (200 combinations). These completely available hourly features are then normalised using the standard score. The last feature to be imputed is therefore the time series with incomplete measurement.

A multivariate imputation is then performed through *IterativeImputer* from sklearn using Bayesian Ridge regression (Pedregosa et al., 2011). The estimated model validity is thus the validation error calculated with the imputed time series as experimental data.

### Weighting adjustment
The feature preparation for weighting methods is slightly different. First, hourly data are aggregated to daily data and only daily features are used. Second, a Principal Component Analysis (PCA) is performed before the weight adjustment to: 1) avoid the problem of multicollinearity, 2) limit the number of auxiliary variables to be used in cell or raking weighting, where both methods require a rational number of variables.

Some primary tests have been launched to decide parameters for different weighting methods, such as the number of used auxiliary variables and used quantiles for discretisation. The adopted parameters are presented in Table 5.

Starting from a base weight of one, the weight of each data point (hourly measurement) is adjusted through three steps:

1. Adjust weights based on the daily data availability,
2. Adjust weights with cell/raking weighting using discretised daily features,
3. Adjust weights to match the population total (complete measured data).

Then the weighted NMBE and CVRMSE are calculated with the following equations (eq. 3 and 4):

$$wNMBE = \frac{\sum_{i=1}^{n} w_i(s_i - m_i)}{\sum_{i=1}^{n} w_i m_i} \times 100 \quad (3)$$

$$wCVRMSE = \sqrt{n} \frac{\sqrt{\sum_{i=1}^{n} w_i(s_i - m_i)^2}}{\sum_{i=1}^{n} w_i m_i} \times 100 \quad (4)$$

where $n$ is the number of data points in the complete measured data (listed in Table 3). The weighted GOF is at last calculated with the eq. 1 to represent the estimated model validity.

*Table 5: Parameters in cell and raking weighting*

|  | CW | RW |
|---|---|---|
| Nb of variables | 5 | 8 |
| Explained variance | 95% | 98% |
| Nb of quantiles | 1-5 | 1-10 |
| Max nb of iteration | - | 2000 |
| Threshold of convergence | - | 1% |

## RESULTS AND DISCUSSIONS

For each substation and the entire sample comprising seven substations, we compared reference errors and error estimates. As all results at both the substation and sample levels exhibit the same trend, we present only sample-level results in Figures 3 and 4.

Figure 3 illustrates how RMSE evolves with varying ratios of missing data using different estimation methods. Figure 4 depicts the bias (median estimate) and variance (95% CI) in the error estimates.

### Unweighted error
The RMSE for unweighted estimation increases almost linearly with missingness. This method shows the highest error under most conditions, except when the missing data exceeds 85%, where raking weighting performs worse. A similar linear trend is observed in the variance evolution (Figure 4). These results indicate that ignoring continuous missing data is not advisable in UBEM validation. Even a small amount of missing data (e.g., 20%) can render the estimated error unrepresentative of the model's overall validity under various weather conditions.

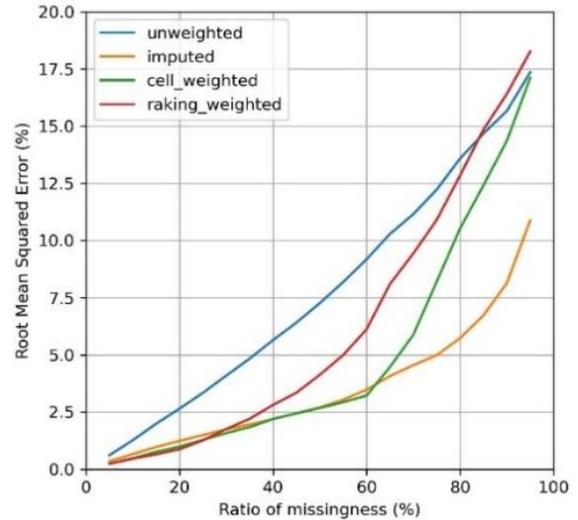

*Figure 3: RMSE calculated for different estimation methods with different ratios of missingness*

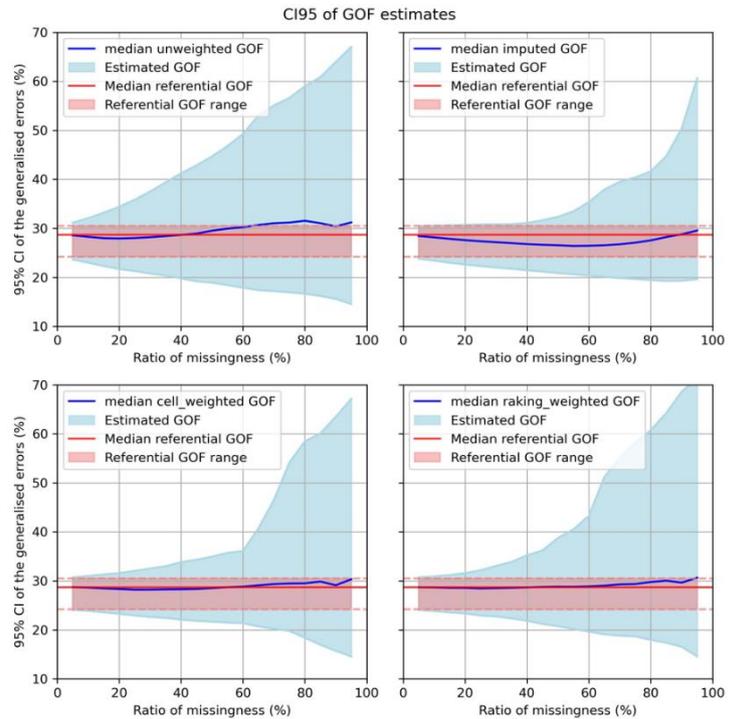

*Figure 4: Median values and 95% CI of error estimates using different estimation methods*

### Imputed error
The estimations of imputation are characterised by their low variance but high bias. The RMSE resulting from imputation remains very low, particularly for large amounts of missing data. However, this RMSE is not always the lowest among the methods: it is slightly higher than the two weighting methods for small missing data ratios (<30%). This can be attributed to the systematic bias observed in Figure 4: the imputation method tends to overestimate model validity, unlike the weighting adjustments. For smaller missingness, such bias caused by imputation is even larger than in unweighted method.

Weighted error

For both weighting methods, trends in median estimates, 95% CI, and RMSE are very similar, with only minor differences. A notable feature in the evolution of the 95% CI is the presence of an "elbow" point. Before this elbow point, the results are characterized by low variance and low bias. However, beyond this point, there is a marked increase in both bias and variance, with the variance becoming the highest among all methods. No significant bias is introduced by missing data until the ratio reaches 60%. Even beyond this point, the bias remains the lowest among all methods, emphasizing the robustness of these methods up to a critical level of missingness.

Usability of incomplete measured data

To evaluate the usability of an incomplete dataset, we define a criterion for the "tolerable missingness": a ratio of missingness is tolerable if 95% of estimated errors are within 10% relative to their reference error. This threshold was selected as a practical criterion to ensure that the majority of estimated errors remain close to the reference model validity, allowing for some level of deviation while still maintaining acceptable accuracy. Based on this criterion, the maximum tolerable missingness for each substation and for the entire sample is shown in Figure 5.

As previously observed, the unadjusted method has the lowest tolerable missingness, while all other estimation methods tolerate a higher ratio of missing data. It is challenging to draw a general conclusion from these results due to the high variability among substations. Specifically, the imputation method exhibits the highest variability, with tolerable ratios of missingness ranging from 17.5% - 37.5% for imputation, 22.5% - 32.5% for cell weighting, and 22.5% - 30% for raking weighting.

Furthermore, the final results are heavily influenced by how we define this criterion, such as the choice between relative or absolute differences and the tolerance level for bias.

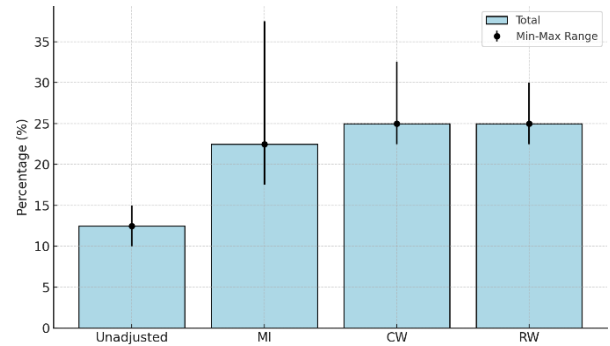

Figure 5: maximum tolerable missingness of the analysed methods across all substations

Methodological implications

Different estimation methods exhibit varying pros and cons. For instance, imputation method typically results in lower variance but higher bias, while weighting methods show lower bias but higher variance and a more consistent performance among different substations.

One limitation of this study is that the detailed implementation of each method can impact the final estimation results. For example, results of weighting methods can depend on the detailed implementation, especially the feature processing stage. PCA applied before weighting reduces dimensionality but can cause an information loss of interactions present in the raw data. Such loss impacts raking weighting more severely than cell weighting, which relies on joint distributions of variables instead of marginal distributions.

All these observations suggest that the choice of estimation method should be made based on the study's objective and context.

Incomplete Data in UBEM Validation

Using incomplete data directly for UBEM validation is not advisable, as indicated by the high errors associated with the unweighted method. This highlights the necessity of employing methods that handle missing data more effectively.

Although estimation bias always exists, adjustment methods from survey research significantly improve the usability of incomplete data, as well as the robustness of UBEM validation using incomplete data.

Considering the bias, variance and stability of estimations, cell weighting appears to outperform the other methods implemented in this study. Its reliance on joint distributions provides evident advantages for handling missing data in UBEM validation.

## CONCLUSION AND PERSPECTIVES

The validation of UBEM is significantly challenged by data scarcity, where the missingness in measured data exacerbates the issue. This scarcity often forces the use of incomplete measured data for UBEM validation, which introduces potential biases, reduces reliability, and can result in large uncertainty in model accuracy assessment. To address these challenges, we employed bias adjustment techniques from survey research, in particular multivariate imputation, cell weighting, and raking weighting.

Our findings indicate that bias adjustment techniques significantly enhance the robustness of UBEM validation by providing more reliable model validity estimates. Among the methods tested, cell weighting exhibited superior performance due to its effective handling of joint distributions of auxiliary variables, making it the preferable approach in our case study. However, the choice of method should be informed by the specific study objectives and dataset characteristics.

Looking forward, several areas for further research emerge. First, a deeper investigation into real-world patterns of missingness in dynamic measured data and their impacts on UBEM validation is needed. Second, identifying which types of buildings or energy uses are particularly sensitive to missing data will help prioritize efforts to reduce data loss impacts and refine validation processes. Third, the exploration of additional bias adjustment techniques, alongside comparative studies, will help establish best practices for managing missing data in UBEM validation. Finally, by making incomplete data usable for validation, we can significantly expand the pool of building samples and validation experiments, which is crucial for defining the validation domain of models—a notable knowledge gap in BEM validation (Ohlsson & Olofsson, 2021).

## ACKNOWLEDGEMENTS

This work was supported by Veolia, by providing meter readings from 34 substations from 2021 to 2023.

## REFERENCES

Ang, Y. Q., Berzolla, Z. M., & Reinhart, C. F. (2020). From concept to application: A review of use cases in urban building energy modeling. *Applied Energy*, *279*, 115738.

Brick, J., & Kalton, G. (1996). *Handling missing data in survey research*.

Gadd, H., & Werner, S. (2014). Achieving low return temperatures from district heating substations. *Applied Energy*, *136*, 59–67.

Garreau, E., Abdelouadoud, Y., Herrera, E., Keilholz, W., Kyriakodis, G. E., Partenay, V., & Riederer, P. (2021). District MOdeller and SIMulator (DIMOSIM) – A dynamic simulation platform based on a bottom-up approach for district and territory energetic assessment. *Energy and Buildings*, *251*.

Kalton, G., & Flores-Cervantes, I. (2003). Weighting Methods. *Journal of Official Statistics*, *19*(2), 81–97.

Lefort, L. (2022). *Methodological developments for the validation of Urban Building Energy Models*.

Mack, C., Su, Z., & Westreich, D. (2018). *Managing Missing Data in Patient Registries Addendum to Registries for Evaluating Patient Outcomes: A User's Guide* (Third Edition).

Morewood, J. (2023). Building energy performance monitoring through the lens of data quality: A review. In *Energy and Buildings* (Vol. 279). Elsevier Ltd.

Ohlsson, K. E. A., & Olofsson, T. (2021). Benchmarking the practice of validation and uncertainty analysis of building energy models. In *Renewable and Sustainable Energy Reviews* (Vol. 142). Elsevier Ltd.

Oraiopoulos, A., & Howard, B. (2022). On the accuracy of Urban Building Energy Modelling. *Renewable and Sustainable Energy Reviews*, *158*, 111976.

Pedregosa, F., Michel, V., Grisel, O., Blondel, M., Prettenhofer, P., Weiss, R., Vanderplas, J., Cournapeau, D., Varoquaux, G., Gramfort, A., Thirion, B., Dubourg, V., Passos, A., Brucher, M., Perrot, M., & Duchesnay, É. (2011). Scikit-learn: Machine Learning in Python. *Journal of Machine Learning Research*, *12*, 2825–2830.

Wang, C., Berthou, T., Duplessis, B., Peirano, E., Schetelat, P., & Riederer, P. (2024). Advancing Urban Building Energy Model Validation: A Comprehensive Multi-period Approach for Dynamic Outputs. *Proceedings of Conference IBPSA France-La Rochelle Oléron-2024*, 631–639.